# Direct evidence of enhanced Ga interdiffusion in InAs vertically aligned free-standing nanowires

J. C. González[1], A. Malachias[2], J. C. de Sousa[1], R-Ribeiro Andrade[1], M. V. B. Moreira[1] and A. G. de Oliveira[1]

[1]Departamento de Física, Instituto de Ciências Exatas, Universidade Federal de Minas Gerais, Postal Code 702, 30123-970, Belo Horizonte, Brazil.

[2]Laboratório Nacional de Luz Síncrotron, C.P. 6192, 13083-970, Campinas, SP, Brazil.

ABSTRACT

We present direct evidence of enhanced Ga interdiffusion in InAs free-standing nanowires grown at moderate temperatures by molecular beam epitaxy on GaAs (111)B. Scanning electron microscopy together with X-ray diffraction measurements in coplanar and grazing incidence geometries show that nominally grown InAs NWs are actually made of $In_{0.86}Ga_{0.14}As$. Unlike typical vapor-liquid-solid growth, these nanowires are formed by diffusion-induced growth combined with strong interdiffusion from substrate material. Based on the experimental results, a simple nanowire growth model accounting for the *Ga* interdiffusion is also presented. This growth model could be generally applicable to the molecular beam heteroepitaxy of III-V nanowires.

III-V free-standing nanowires (NWs) have recently attracted considerable interest because of their potential for optoelectronic applications. NWs have been regarded as ideal systems for an understanding of the role of dimensionality and size in optical, electrical, and mechanical properties of nano-objects. Several research groups [1-7] have made progress leading to precisely controlled synthesis of III-V NWs. However, despite such progress, the growth mechanism that governs the formation of these nanostructures has not been thoroughly explored, nor is it fundamentally understood. The theoretical models of the formation of NWs can be divided into two large groups [8]. In the first group are the models based on the classical "vapor–liquid–solid" (VLS) mechanism. First suggested by Wagner and Ellis [9], the VLS is the most commonly accepted model of NWs growth. In the VLS model, a liquid droplet of a catalyst is formed at high temperatures and it is assumed that the precursors that fall into the droplet from the gas phase firstly dissolve in the droplet and then crystallize at the liquid solid interface under the droplet. The droplet then moves upward at a rate equal to the vertical rate of growth of the NWs. The VLS mechanism, initially suggested to describe the growth of micrometer-sized crystals [9], was later used to explain the formation of different types of NWs [10, 11]. In the second group of models, the growth of a NW is stimulated by surface diffusion of precursor's adatoms to the NW top. In this case the growth of a NW is not only due to the direct impingement of gas phase atoms onto the catalyst droplet, but also to the diffusion flux of adatoms. These adatoms, originated from the gas phase, reach the substrate surface and migrate to the NW sidewalls towards the catalyst droplet. The classical VLS mechanism is dominant in relatively thick NWs, while the diffusion induced growth is dominant for thin NWs. V. G. Dubrovskii et al. [8] have reviewed the different theoretical models of the formation of NWs and developed a model of diffusion-induced growth of NWs applicable to a large variety of technologies of growth which accounts for the surface diffusion of adatoms. Nevertheless, most of the experimental works on this area have focused on the NWs length as a function of the NWs radius [8, 12-15] or on the dependence of the growth rate on growth parameters such as the distribution of the catalyst [14], temperature of growth [12-15], migration length of adatoms [14] and flux of precursors [12-15], etc. Despite such theoretical and experimental studies, we have not been able to find reports on structural and/or chemical modifications of the NWs due to the interdiffusion contribution of adatoms coming from the top monolayers of the substrate. This interdiffusion process has been proven to strongly modify the physical properties of other free-standing nanostructures such as self-assembled quantum dots [16, 17] and rings [18].

In this work we present experimental results of scanning electron microscopy and X-ray diffraction (XRD) indicating that nominally grown InAs free-standing NWs, on GaAs, are actually $In_{0.86}Ga_{0.14}As$ NWs. The GaAs molar fraction in the NWs has been established by measuring the longitudinal and the radial lattice parameters of the NWs. The formation of the alloy is then attributed to the diffusion of *Ga* adatoms from the top monolayers of the substrate towards the sidewalls of the NWs and the catalyst droplet. Based on the experimental results, a simple nanowire growth model accounting for the *Ga* and *In* diffusion is also presented. This growth model could be generally applicable to the molecular beam heteroepitaxy of III-V nanowires.

Vertically aligned free-standing InAs NWs were grown by molecular beam epitaxy (MBE) using 5 nm colloidal gold nanoparticles as catalyst. A GaAs (111)B substrate was drop coated with the catalyst and then introduced and degassed in a ultra high vacuum (UHV) chamber at 400 °C for 2 hours. Following this, the substrate was transferred to the UHV growth chamber and deoxidized at 620 °C under an $As_4$ flux for 20 min. The substrate temperature was then lowered to 510 °C and the *In* cell was opened. The beam equivalent pressures for the *In* and $As_4$ used were $3.1\times10^{-7}$ Torr and $5.8\times10^{-5}$ Torr, respectively. The *Ga* cell was kept cold and closed during the whole growth procedure. After 20 minutes of growth the *In* cell was closed, the sample cooled to room temperature and then removed from the MBE system.

Scanning electron microscopy measurements were carried out on a JSM 6330F field emission microscope in order to study the morphology of the sample. X-ray diffraction experiments were carried out on the XRD2 beamline of the Brazilian National Light Synchrotron Laboratory at 10KeV x-ray energy. This beamline is equipped with a 4+2-circle Huber diffractometer and a position sensitive detector (PSD) that integrates a solid angle of 1.5º. Grazing incidence diffraction (GID) and coplanar diffraction measurements were used to investigate the in-plane and out-of-plane (surface perpendicular) crystal structure of the NWs, respectively.

Figure 1 shows a scanning electron micrograph of the NWs. The SEM measurements show that the NWs have a mean diameter of (46±13) nm and most of them are vertically aligned with respect to the substrate. The inset of Figure 1 shows a top view SEM image of a single NW. According to these images the nanowires are composed of six vertical $\{\bar{1}10\}$ facets and a top (111)B plane. Large islands or grains, forming a rough surface, are clearly visible between the NWs. The mean height of the islands was found to be about 400 nm, while the tallest nanowires have approximately 2000 nm. The estimated mean growth rates of the islands and nanowires were 1.2 µm/h and 3 µm/h, respectively.

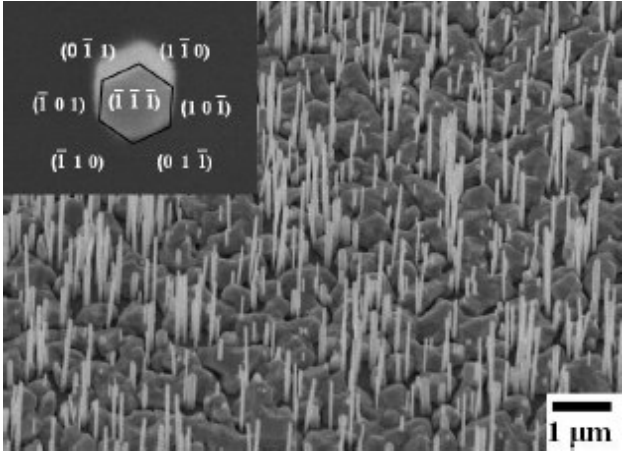

**Figure 1**. 30°-perspective view of a SEM image of the vertically oriented free-standing $In_{0.86}Ga_{0.14}As$ NWs. Large InAs islands are also observed in between the nanowires. The inset shows a top view of a hexagonal NW with the indication of the lateral facets.

Figure 2(a) shows three XRD radial scans corresponding to the $(2\bar{2}0)$ and $(\bar{2}4\bar{2})$ in-plane reflections, as well as the $(\bar{3}3\bar{3})$ out-of-plane reflection. A sketch of the directions of x-ray measurements is shown in Fig. 2(b). All scans span from the InAs to the GaAs reciprocal space positions. In the graphs the momentum transfer (q) axis was directly converted to the local lattice parameter to allow a direct comparison of in-plane and out-of-plane measurements. The $(2\bar{2}0)$ and $(\bar{2}4\bar{2})$ reflections directly probe the lattice parameter in directions perpendicular to the $\{1\bar{1}0\}$ facets of the NWs and along a diagonal of the hexagonal cross-section of the NWs, respectively. On the other hand, the $(\bar{3}3\bar{3})$ reflection probes the longitudinal lattice parameter of the NWs. In all scans strong diffraction peaks are observed at the bulk InAs and GaAs reciprocal space positions corresponding to lattice parameters of

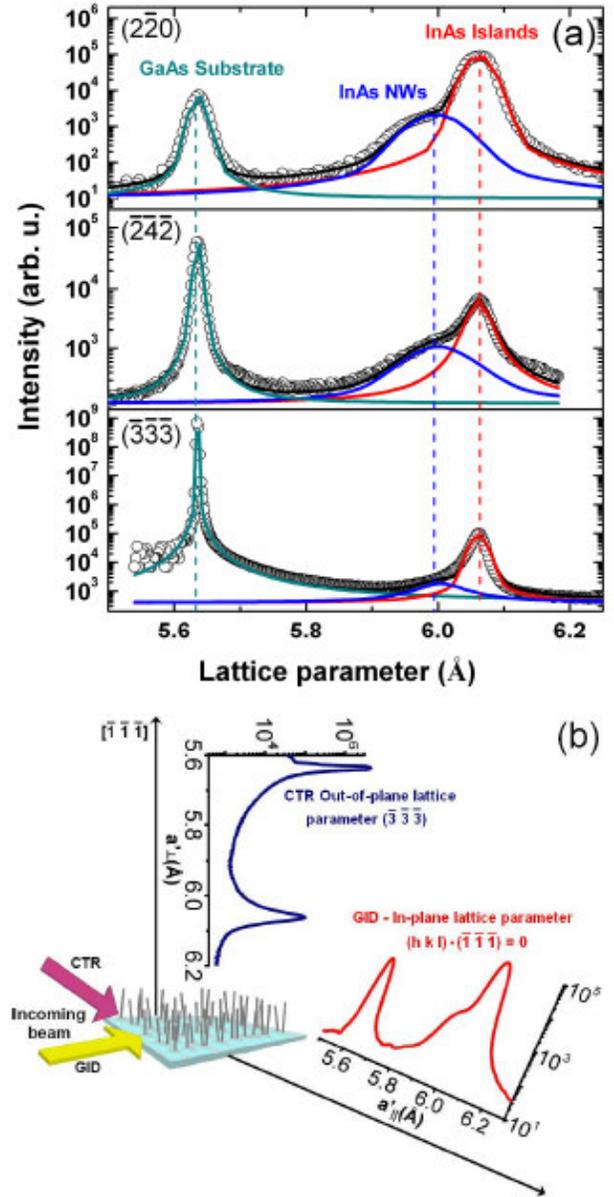

**Figure 2.** (a) Radial scans of the $(2\bar{2}0)$ and $(\bar{2}4\bar{2})$ and $(\bar{3}3\bar{3})$ reflections of the samples. Note that the intensity hump in the left side of the InAs peak (islands) appears in the same position in the three scans. (b) Schematics of the geometry of the experiments for the coplanar and GID configurations.

6.058Å and 5.653 Å, respectively [19]. While the InAs peak can be uniquely ascribed to a fully relaxed pure InAs structure, an intensity hump can be clearly observed on the left side (smaller lattice parameter) of the InAs position in both $(2\bar{2}0)$ and $(\bar{2}4\bar{2})$ radial scans at the local lattice parameter of 5.999(7) Å. Such a hump is also observed at the same equivalent reciprocal space position of the out-of-plane $(\bar{3}\bar{3}3)$ reflection.

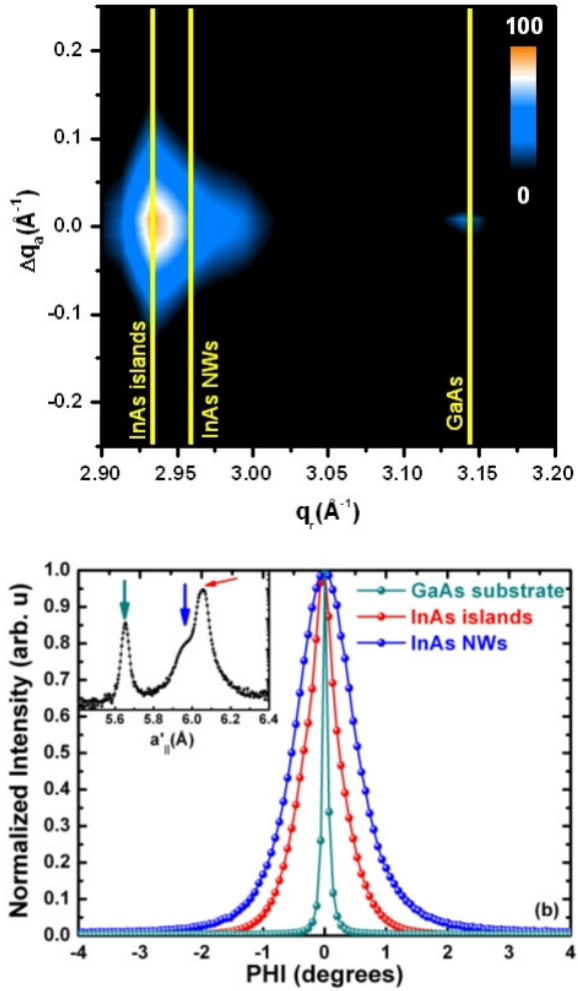

**Figure 3.** (a) Reciprocal space map of the $(2\bar{2}0)$ reflection. (b) Normalized angular scans taken at the position of the substrate, NWs and islands. The lines connecting the dots are to guides the eyes. Insert: radial scan of the $(2\bar{2}0)$ reflection with the positions of the angular cuts indicated with arrows.

In order to clarify which structure, NWs or islands, corresponds to the bulk InAs peak and which to hump, angular (transversal) scans were also measured along the $(2\bar{2}0)$ radial scan.

Figure 3(a) shows a reciprocal space map of the $(2\bar{2}0)$ reflection. Three normalized angular scans taken at the position of the substrate peak, InAs peak and InAs hump were extracted from the reciprocal space map and are shown in Figure 3(b). These scans exhibit very different angular widths. The width $\omega_a$ of the angular scans may have two origins [20]. Firstly, the limited size $D = 2\pi/\Delta q_S$ of crystallites may broaden the diffraction peak with a contribution $\omega_S(q) = \Delta q_S / q$, where $\Delta q_S$ is the constant peak broadening in reciprocal space due to the finite-size of the crystallites and $q$ is the momentum transfer. Secondly, the mosaic spread of crystallites add a component $M$ to the angular peak width that can be taken into account by the relation $\omega_a^2 = \omega_S^2 + M^2 + \omega_i^2$, where $\omega_i$ accounts for the instrumental broadening in the experiments taken here to be equal to the angular width of the substrate peak. Since the InAs hump is very close to the InAs peak, we have assumed that the crystalline structures in the sample have negligible strain or are fully unstrained. In fact, when taking strain broadening into consideration, the size of the crystallites obtained from those calculations will alter by less than 1%.

In order to evaluate the size of the crystallites and the mosaic spread, the radial scans must be also analyzed. The width of the x-ray peaks in the radial direction $\omega_r$ is similarly given by the strain broadening $\omega_{str}$, the crystallite size broadening $\omega_s$ and the instrumental broadening ($\omega_i$), i.e., $\omega_r^2 = \omega_{str}^2 + \omega_s^2 + \omega_i^2$. Having ruled out the presence of strain in our sample, the angular width of the radial scans $\omega_r$ is roughly given by the $\omega_s$ component. Fitting both, the radial and angular scans with Voigt functions $\omega_r$ and $\omega_a$ were extracted. Using the previous relationships we have found that the mean lateral size of the objects diffracting in the position of the InAs peak is (164±4) nm with a mosaic spread of (0.57±0.02) degrees. The corresponding values for the InAs hump are (39±4) nm and (0.65±0.02) degrees, respectively. These dimensions are similar to the lateral sizes of the InAs islands and NWs as found by scanning electron microscopy. The mosaic spread of the InAs hump is slightly above the value obtained for the islands, indicating a small tilt distribution of the NWs as observed in Figure 1.

Another piece of evidence that points to the relation between the diffraction intensity observed at the hump and the NWs is provided by the scattering profile measured by the PSD in GID measurements along the surface perpendicular direction. Figure 4 shows the scattering profile along $\alpha_f$ measured by the PSD in a range of 1.5º for the GaAs, InAs and hump reciprocal space positions at the $(2\bar{2}0)$ reflection. For the GaAs position the scattering peak is observed exactly at the critical angle $\alpha_c$ ($\alpha_c = 0.258º$ for GaAs at E = 10KeV). The scattering profile measured at the InAs peak position has a similar shape to the one measured at the GaAs position, but with the scattering peak at a slightly lower exit angle ($\alpha_f = 0.188º$). Both the GaAs and InAs scattering profiles are typical examples of the exit angle intensity distribution in grazing incidence diffraction of real surfaces and can be interpreted within the distorted wave born approximation (DWBA) [21, 22]. Therefore, the observed InAs peak in the $(2\bar{2}0)$ radial scans is originated by a structure that is slightly above the GaAs surface, i.e. the effective rough film created by the InAs islands. On the other hand, the scattering profile acquired at the position of the InAs intensity hump presents a completely different shape. This profile is consistent with diffraction from the bulk (Born approximation) or an extremely rough film [21, 22] with no contribution from the surface Fresnel reflectivity. Hence, the absence of a scattering peak near the InAs critical angle ($\alpha_c = 0.188°$ at E = 10KeV) is a clear fingerprint of direct scattering from objects that are spatially apart from the substrate.

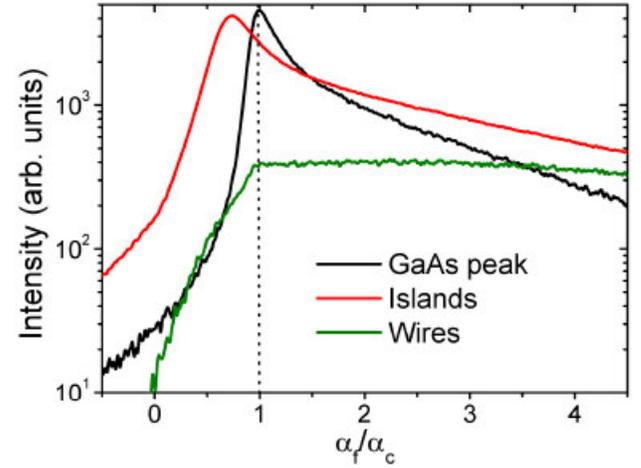

**Figure 4.** Scattering profiles along $\alpha_f$ obtained by the position sensitive detector at the substrate, islands and nanowires peaks for the $(2\bar{2}0)$ reflection. The scattering from wires is compatible with a three-dimensional structure and does not exhibit the characteristic surface peak as observed for the islands and substrate.

Based on the above results and discussions we can conclude that the intensity hump observed in the radial scans of Figure 2(a) corresponds to the x-ray diffraction from the InAs NWs, while the InAs peak is due to the diffraction from the InAs islands. However, the origin of the nanowires lattice parameter shift is still unclear. In the following we will discuss this aspect.

The intensity hump observed in the in-plane radial scans could be, in principle, due to an in-plane compressive biaxial strain in the NWs caused by the lattice mismatch, of approximately 7%, between InAs and GaAs. In fact, recent theoretical and experimental works have shown evidence that below a critical diameter it is possible to obtain coherent growth of strained NWs on lattice-mismatched substrates [24, 25]. These works indicate that InAs NWs grown on GaAs substrates have a critical diameter of approximately 40 nm, just like our NWs. However, in-plane biaxially strained InAs, NWs must have an out-of-plane lattice parameter expansion, i.e. the intensity hump in the $(\bar{3}\bar{3}3)$ reflection is expected to be observed on the right side of the $(\bar{3}\bar{3}3)$ InAs peak. In addition, InAs grown on GaAs are known to relax after just a few monolayers of deposition [26, 27]. Therefore, the intensity hump observed at the same lattice parameter position for all radial scans cannot be assigned to strain effects.

An alternative interpretation of the intensity hump observed at 5.999(7) Å in the $(2\bar{2}0)$, $(\bar{2}4\bar{2})$ and $(\bar{3}\bar{3}3)$ radial scans is to assume a possible incorporation of *Au* from the catalyst droplet into the nanowires. We have not found reports on Au-In-As compounds, but there are several reported *Au-In* and *Au-As* crystalline compounds [28]. However, we have carried out long radial scans in GID and coplanar geometries and no extra peak indicating the presence of those compounds was found. Resonant GID experiments were also carried out at the E=11KeV absorption edge of Au looking for a substantial (more than 3%) incorporation of Au into the InAs crystalline lattice, with negative results.

Finally, a more plausible explanation for the intensity hump is the formation of an InGaAs alloy in the nanowires during growth. In this case the NWs should have a cubic crystal structure with the same lattice parameter along the radial and longitudinal directions, just as was observed experimentally. The observed lattice parameter corresponds to an unstrained $In_xGa_{1-x}As$ alloy with an average InAs molar fraction $x = (0.86\pm0.02)$. The formation of the $In_xGa_{1-x}As$ alloy will be explained in the next paragraphs, considering the NWs growth model developed by Dubrovskii et al. [29].

In the classical VLS model the growth of an InAs NW, of length $L$ and radius $R$, on (111)B GaAs is caused by the direct impingement of *In* atoms on the surface of the catalyst droplet that lies on top of the nanowire. However, a large part of the atomic *In* flux directly impinges the substrate surface, creating *In* adatoms that can diffuse to the top of the NW contributing to the growth rate [8, 14, 15]. This diffusion-induced growth process is illustrated in Figure 5. The steady-state growth rate of the NW can be expressed as $V_{NW} = V_{VLS} + V_{diff}$, where $V_{VLS}$ is the growth rate due to the direct impingement of *In* atoms on the catalyst drop as explained by the VLS growth model [9] and $V_{diff}$ is the diffusion-induced growth rate due to the diffusion flux of *In* adatoms from the substrate surface to the NW. At this point we will generalize the last idea by considering the existence of an equilibrium concentration of *Ga* adatoms, supplied from the top monolayers of the GaAs substrate, which can also diffuse towards the NW. Therefore, the $V_{diff}$ should be rewritten as $V_{diff} = V_{diff}^{In} + V_{diff}^{Ga}$, where $V_{diff}^{In}$ and $V_{diff}^{Ga}$ are the separated contributions to the growth rate due to the diffusion flux of *In* and *Ga* adatoms.

Comparing the growth rate of thick (d ~120 nm) and thin (d~30 nm) NWs in our sample, we have found that for a typical 40 nm-thick NW the diffusion-induced growth regime is dominant over the classical VLS regime [29]. This conclusion is also in agreement with the results of ref. [29], which points out that the classical VLS growth mode is dominant for III-V NWs with a diameter larger than 100 nm. Therefore, the VLS growth rate can be neglected and the InAs molar fraction $x$ in the average NW can be approximated as follows:

$$x = V_{diff}^{In} / V_{diff} = V_{diff}^{In} / (V_{diff}^{In} + V_{diff}^{Ga}) \quad (1)$$

Both *In* and *Ga* diffusion growth rates are proportional to the diffusion flux of *In* and *Ga* adatoms from the substrate surface to the NW top. Following Dubrovskii et al. [29], the diffusion flux $j_{diff}(L)$, for each element, can be expressed by:

$$j_{diff}(L) = j_{diff}(0) \cdot \left( \frac{1}{\cosh(L/\lambda_f)} - c \cdot (\xi+1) \cdot \tan(L/\lambda_f) \right) (2)$$

where $j_{diff}(0)$ is the diffusion flux of adatoms from the substrate surface to the NW base, $\lambda_f$ is the diffusion length of the element (*Ga* or *In*) in the sidewalls of the NW, $c$ is a coefficient related with the equilibrium adatom coverage on the sidewalls of the NW, and $\xi$ is the supersaturation in the catalyst drop. Considering that under typical MBE conditions the second term to the right of Eq. (1) is much smaller than the first term, and can be neglected [29], and $\lambda_f$ to be of similar magnitude for *In* and *Ga* adatoms [30], then:

$$1/x = 1 + j_{diff}^{Ga}(0) / j_{diff}^{In}(0) \quad (3)$$

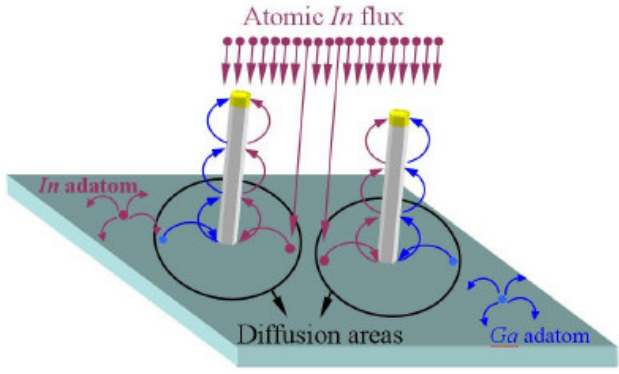

**Figure 5.** Schematic illustration of the diffusion-induced growth process of InGaAs nanowires. The *In* and *Ga* adatoms inside the diffusion area of each nanowire will diffuse towards the NW and contribute to its growth. The adatoms outside those diffusion areas will contribute to the formation of islands.

Following ref. [29], the diffusion flux of adatoms from the substrate surface to the NW base can be calculated as $j_{diff}(0) = l_s \pi R \sigma N_{eq}/t_s$, where $l_s$ is the length of adatom diffusion jump on the main surface, $t_s$ is the characteristic time between jumps, $\sigma$ is the adatom supersaturation, and $N_{eq}$ is the equilibrium adatom concentration on the substrate surface. The length of adatom diffusion jump and the characteristic time between jumps for each element should be very similar. Furthermore, the supersaturation should be of the order of unity [31]. The InAs molar fraction can be calculated then from

$$1/x = 1 + N_{eq}^{Ga}/N_{eq}^{In} \tag{4}$$

In our case, the ratio between the *Ga* and *In* equilibrium adatom concentrations on the substrate surface should be approximately 0.16, i.e. the concentration of *In* adatoms is nearly six times larger than the concentration of *Ga* adatoms. Direct measurement of adatom concentrations is extremely difficult, especially under growth conditions. However, both concentrations are dependent on the growth temperature, the primary *In* atom flux and somehow on the $As_4$ flux as well [32]. This gives growers the added flexibility of tuning three parameters independently to control the chemical composition of the $In_xGa_{1-x}As$ NWs and, therefore, their electronic and optical properties.

In this work we have studied, by using scanning electron microscopy and x-ray diffraction techniques, the morphology, crystal structure and chemical composition of free-standing InAs NWs grown on a (111)B GaAs substrate by molecular beam epitaxy. We have shown direct evidence that *Ga* adatoms from the top monolayers of the GaAs substrate surface play an important role in the growth mechanism of the InAs NWs and significantly modify the chemical composition of these elements. The incorporation of *Ga* in the NWs is of the order of 14%. We have also shown that a generalization of the diffusion-induced growth model of free-standing nanowires, to include the diffusion of *Ga* adatoms from the substrate to the NWs, can satisfactorily explain our results. This growth model could be generally applicable to the molecular beam heteroepitaxy of III-V nanowires.

ACKNOWLEDGMENT

We would like to thank the Brazilian National Light Synchrotron Laboratory, CNPq, CAPES and FAPEMIG for the financial support of this work.